\pdfoutput=1

\documentclass{iacrtrans}

\usepackage{xcolor}
\usepackage{listings}
\usepackage{float}
\usepackage{rotating}
\usepackage{caption}
\usepackage[utf8]{inputenc}
\usepackage{csquotes}
\usepackage{amsmath, amsthm}
\usepackage{tikz}
\usepackage{pgfplots}
\pgfplotsset{compat=1.18}
\usepackage{mathrsfs}
\usepackage{amssymb}
\usepackage{threeparttable}
\usepackage{algorithm}
\usepackage[noend]{algpseudocode}
\usepackage{url}
\usepackage[normalem]{ulem} % used for strikethrough

\usepackage{graphicx}

\usepackage{placeins}

\usepackage{pifont}

\usepackage{ifthen}
\usepackage{hyperref}
\usepackage[capitalize,nameinlink]{cleveref}
\usetikzlibrary{matrix}
\usetikzlibrary{arrows.meta}
\newcommand{\ultratiny}{\@setfontsize\ultratiny{4}{5}}%

\newcommand{\F}{\ensuremath{\mathbb{F}}\xspace}

\newcommand{\Z}{\ensuremath{\mathbb{Z}}\xspace}

\newcommand{\shared}[1]{\ensuremath{[{#1}]}}

\newcommand*{\numero}{n\kern-.1em \raise.7ex\vbox{\hbox{\tiny \ensuremath{\circ}}\kern.5pt}}

\usepackage{enumitem,booktabs}
\usepackage[referable]{threeparttablex}
\renewlist{tablenotes}{enumerate}{1}
\makeatletter
\setlist[tablenotes]{label=\tnote{\alph*},ref=\alph*,itemsep=\z@,topsep=\z@skip,partopsep=\z@skip,parsep=\z@,itemindent=\z@,labelindent=\tabcolsep,labelsep=.2em,leftmargin=*,align=left,before={\footnotesize}}
\makeatother

\usetikzlibrary{arrows}
\usetikzlibrary{decorations.markings}
\tikzset{XOR/.style={draw,circle,append after command={
        [shorten >=\pgflinewidth, shorten <=\pgflinewidth,]
        (\tikzlastnode.north) edge (\tikzlastnode.south)
        (\tikzlastnode.east) edge (\tikzlastnode.west)
        }
    }
}

\tikzstyle{int}=[draw, fill=blue!20, minimum size=2em]
\tikzstyle{init} = [pin edge={to-,thin,black}]

\usetikzlibrary{decorations.pathreplacing,calc}

\usetikzlibrary{shapes,arrows,fit,calc,positioning,automata,decorations.pathreplacing,patterns,decorations.pathmorphing}
\tikzset{XOR/.style={draw,circle,scale=1,append after command={
        [shorten >=\pgflinewidth, shorten <=\pgflinewidth,]
        (\tikzlastnode.north) edge (\tikzlastnode.south)
        (\tikzlastnode.east) edge (\tikzlastnode.west)
        }
    }
}
\tikzset{MUL/.style={draw,circle,scale=1,append after command={
        [shorten >=\pgflinewidth, shorten <=\pgflinewidth,]
        (\tikzlastnode.north west) edge (\tikzlastnode.south east)
        (\tikzlastnode.south west) edge (\tikzlastnode.north east)
        }
    }
}
\tikzset{buswidth/.style={draw=none,circle,scale=1,append after command={
        [shorten >=\pgflinewidth, shorten <=\pgflinewidth,]
        (\tikzlastnode.south west) edge (\tikzlastnode.north east)
        }
    }
}
\tikzset{enc/.style=   {draw=black,fill opacity=.3},
         encdec/.style={draw=black,fill opacity=1}
}

\usepackage{booktabs}
\usepackage{todonotes}
\usetikzlibrary{matrix}
\usetikzlibrary{shapes,fit}
\usetikzlibrary{patterns}

\usepackage{footnote}
\usepackage{comment}

\usepackage{xspace}

\usepackage{cleveref}
\floatstyle{boxed}
\newfloat{scheme}{ht}{sc}
\floatname{scheme}{Scheme}
\Crefname{scheme}{Scheme}{Schemes}

\newfloat{functionality}{ht}{fc}
\floatname{functionality}{Functionality}
\Crefname{functionality}{Functionality}{Functionalities}
\usepackage{xspace}
\usepackage{multirow}

\usepackage[n, advantage , operators ,
sets ,
adversary ,
landau ,
probability ,
notions ,
logic ,
ff ,
mm,
primitives ,
events ,
complexity ,
asymptotics ,
keys
]{cryptocode}

\begin{document}

\title{Large-Scale MPC: Scaling Private Iris Code Uniqueness Checks to Millions of Users}

\author{
    Remco Bloemen \inst{2}\and
    Bryan Gillespie \inst{3}\and
    Daniel Kales \inst{1}\and
    Philipp Sippl \inst{2}\and
    Roman Walch \inst{1}
}
\institute{
    TACEO, Graz, Austria \\
    \email{lastname@taceo.io}\and
    Worldcoin Foundation\and
    Inversed Tech\\
    \email{bryan@inversed.tech}
}

\maketitle

\begin{abstract}
    In this work we tackle privacy concerns in biometric verification systems that typically require server-side processing of sensitive data (e.g., fingerprints and Iris Codes).
    Concretely, we design a solution that allows us to query whether a given Iris Code is similar to one contained in a given database, while all queries and datasets are being protected using secure multiparty computation (MPC).
    Addressing the substantial performance demands of operational systems like World ID and aid distributions by the Red Cross, we propose new protocols to improve performance by more than three orders of magnitude compared to the recent state-of-the-art system Janus (S\&P 24).
    Our final protocol can achieve a throughput of over 690 thousand Iris Code comparisons per second on a single CPU core, while protecting the privacy of both the query and database Iris Codes.
    Furthermore, using Nvidia NCCL we implement the whole protocol on GPUs while letting GPUs directly access the network interface. Thus we are able to avoid the costly data transfer between GPUs and CPUs, allowing us to achieve a throughput of 4.29 billion Iris Code comparisons per second in a 3-party MPC setting, where each party has access to 8 H100 GPUs.
    This GPU implementation achieves the performance requirements set by the Worldcoin foundation and will thus be used in their deployed World ID infrastructure.
    
\end{abstract}

\section{Introduction}

Decentralized blockchains very fundamentally rely on the existence of a sybil-resistant mechanism for their consensus.
While the mechanisms applied in this context (e.g. Proof-of-Stake or Proof-of-Work) make it very hard to accumulate a large share of representative power, they do not enforce that each participant owns \textit{exactly} one share of equal size.
However, there are many practical applications, in which it is of utmost importance to ensure that users can only sign up \textit{exactly} once for a given service.

For example, Worldcoin's World ID\footnote{\url{https://worldcoin.org/world-id}} is a privacy-preserving  \textit{Proof-of-Personhood}, which requires that only real humans are able to sign up exactly once and are then able to reuse this proof for other applications.
Another very relevant example is aid distribution by, e.g., the Red Cross \cite{JanusSP24}.
While everyone should be able to apply for the aid program, people should be prevented from signing up multiple times, since available resources are usually scarce and should be used to help as many people as possible.

In both examples, the \emph{uniqueness} property is enforced by storing biometric information, such as so-called Iris Codes derived from the human iris, in a database on a central server.
When new users sign up, their biometric information is captured and some sort of liveness check is executed to ensure it is an authentic scan; afterwards their biometric information is sent to this server which checks whether it is close to any other sample in the database.
If no match was found, the user is allowed to sign up and their biometric information is added to the database.

While this simple protocol ensures uniqueness, it requires a single server to collect and store information derived from biometric data which may pose privacy concerns to some.
Leakage of such a database could become even more critical when this biometric database is linked to other datasets.
Furthermore, in such cases the database holders may potentially misuse the dataset to, e.g., discriminate against specific ethnicities,\footnote{\url{https://www.hrw.org/news/2022/03/30/new-evidence-biometric-data-systems-imperil-afghans}} and it becomes a prime target for attackers who might try to steal the data.

In this paper we propose an efficient and scalable protocol to ensure uniqueness of biometric information, concretely Iris Codes, in a distributed database.
The database, thereby, is protected by secure multiparty computation (MPC), a cryptographic technique which allows multiple parties to compute functions on private data without revealing their private inputs to each other.
Concretely, the database is split amongst multiple parties, such that no party on their own learns anything about the contained biometric information.
Furthermore, new Iris Codes are also shared using MPC, protecting their privacy as well.

While MPC solves the privacy issues of the uniqueness service above, indiscriminately applying it to any use case comes with a significant performance cost (see, e.g., \cite{JanusSP24}).
Even more critical for real-world deployment are the performance requirements of large-scale systems, such as the World ID infrastructure which we focus on, where the overall system needs to scale to a database of tens of millions of Iris Codes while supporting a throughput of 10 queries per second.
Looking at these numbers we can already draw two conclusions:
(i) None of the related work comes close to achieving these requirements (see also next section).
Hence, we have to come up with more efficient approaches and protocols, while also leveraging all available hardware, such as GPU's.
(ii) The final solution will inevitably involve having a cluster of computing nodes for each MPC party.
Hence, the main goal of this work is making the protocol as efficient as possible to reduce the required number and performance requirements of cluster nodes and, therefore, the overall cost.

\subsection{Related Work}

Many privacy-preserving solutions for authentication using biometric information have been introduced in the literature \cite{DBLP:journals/spm/BringerCP13,DBLP:journals/eswa/ImamverdiyevTK13,DBLP:conf/ccs/JuelsW99,DBLP:journals/mta/KumarPR20,DBLP:journals/iet-bmt/Tams16,DBLP:conf/infocom/YuanY13,DBLP:journals/access/ZhuZXLH18}, with the recently published Janus \cite{JanusSP24} (S\&P 2024) being the most similar one to our solution.
In Janus, the authors propose privacy-preserving protocols to prevent recipients from registering to an aid distribution process multiple times by using biometric information.
In essence, their deduplication protocol is very similar to the one employed by Worldcoin, i.e., using similarity measurements such as the hamming distance to determine whether biometric data of a new participant matches that of already registered ones.
However, while we use a 3-party MPC solution (see later sections) to protect new and existing biometric data, they compare three different approaches, based on 2-party garbled circuits \cite{DBLP:conf/focs/Yao86}, somewhat homomorphic encryption (SHE) \cite{DBLP:phd/us/Gentry09}, and trusted execution environments (TEEs).
Though they acknowledge that the security guarantees of TEEs are broken regularly by software- and hardware-based side channel attacks (see, e.g., \cite{DBLP:conf/uss/Lipp0G0HFHMKGYH18, DBLP:conf/uss/BulckMWGKPSWYS18} and \url{https://sgx.fail/}), their single-threaded TEE solution outperforms SHE and MPC and achieves deduplication of a database with 8000 users in 50 ms.
They additionally show that their SHE solution outperforms their MPC-based one, while requiring a higher deduplication runtime of $40$ s.\ for the 8000 user database.

As we demonstrate in \Cref{sec:benchmarks}, the solution described in this work significantly outperforms those explored in Janus by several orders of magnitude. Our protocol (even without GPU acceleration) can achieve single-threaded throughput of 690 thousand Iris Codes per second, while supporting Iris Codes with larger precision.
When computing parties have access to 8 H100 GPUs, each connected by fast network connections, we show how to improve the performance to 4.29 billion Iris Code comparisons per second.
Consequently, we believe our protocol can be used to significantly speed up biometric deduplication checks for aid distribution as well.
Furthermore, as described in \cite{JanusSP24}, the non-collusion assumption between the three computing nodes in our design can easily be achieved in the context of aid distribution by the International Committee of the Red Cross (ICRC).

Only a handful of papers consider GPU acceleration for MPC computations.
Early publications focused on implementation of basic and non-specialized protocols \cite{DBLP:conf/acns/FrederiksenN13,DBLP:conf/acsac/HustedMSG13}, while
more recent work has developed techniques for machine learning training and inference \cite{DBLP:conf/uss/WatsonWP22,DBLP:conf/uss/MishraLSZP20,DBLP:conf/sp/TanKTW21,jiang2024spin}.
The computationally most intensive part of our proposed protocol is computing dot-products, so our protocol immediately benefits from GPUs in a similar fashion as machine learning use cases.
However, we go one step further and implement hardware acceleration of a complete MPC protocol with multiple tens of communication rounds by giving the GPUs direct access to the network interface.
This allows the design to avoid costly data transfers between GPUs and CPUs, meaning the GPU acceleration provides performance benefits throughout the entire protocol execution.

\section{MPC Background}

Secure multiparty computation (MPC) allows multiple parties to compute a function on their combined private input data without leaking these inputs to each other.
In this paper we focus on secret sharing based MPC protocols \cite{DBLP:journals/cacm/Shamir79,DBLP:conf/crypto/DamgardPSZ12,DBLP:conf/ccs/MohasselR18}.
In these protocols, private data is shared amongst $n$ parties, such that each party holds a share which on its own does not leak any information about the original dataset.
Given enough shares, however, the parties can reconstruct the data.
Furthermore, the parties can use the shares to compute algebraic functions, which in turn yield shares of the result.
\emph{Linear} operations, such as adding shares or multiplying them with constants, can directly be computed on the shares, whereas multiplications of two shared values require the parties to exchange some form of randomized information.
Consequently, the more multiplications that are present in the circuit one wants to compute, the more data that needs to be exchanged by the parties.  As a result, the number of multiplications in the circuit is a large part of the cost metric of MPC protocols, and should be reduced for efficiency.

Many different flavours of MPC protocols exist, such as protocols based on additive secret sharing  (e.g., SPDZ \cite{DBLP:conf/crypto/DamgardPSZ12}), replicated secret sharing (e.g., ABY3 \cite{DBLP:conf/ccs/MohasselR18}, Swift \cite{DBLP:conf/uss/KotiPPS21}, Fantastic Four \cite{DBLP:conf/uss/Dalskov0K21}), Shamir's secret sharing \cite{DBLP:journals/cacm/Shamir79}, and Yao's garbled circuits \cite{DBLP:conf/focs/Yao86}.
These protocols differ in the general sharing techniques, supported number of computing parties, tolerated number of malicious parties, and general security model.

Due to the high performance requirements of the World ID infrastructure, we focus in this work on protocols which require an \textit{honest majority} among the parties (see \Cref{sec:hd}).
Furthermore, similar to related work (e.g., \cite{JanusSP24}), we focus on the \emph{semi-honest} security model.\footnote{All protocols in this paper can be adapted to provide malicious security as well. See the report contained in \url{https://github.com/TaceoLabs/worldcoin-experiments} for some preliminary experiments.}
Finally, since MPC protocols usually suffer from high overall communication costs, and since deployments are likely to be running within server farms with low-latency network connections, our efforts focus on reducing the total communication of the protocol, potentially accepting a higher number of sequential communication rounds to do so when such a trade-off exists.

\paragraph{Notation.}
Throughout the paper, $\Z_t$ denotes the integers mod $t$, and a secret sharing of a value $x\in\Z_t$ is written as $\shared{x}$.
Additionally, we write $\vec{x} \odot \vec{y}$ for the element-wise multiplication (Hadamard product) of two vectors $\vec{x},\vec{y}\in\Z_t^\ell$.

\section{The Iris Code Membership Protocol deployed by World ID} \label{sec:plain}

In this work, we focus on the uniqueness services deployed by Worldcoin for their World ID infrastructure, which is why we use Iris Codes \cite{DBLP:conf/icip/Daugman02} as the biometric data of choice.
Iris Codes are derived from images of human eyes and are considered as a highly accurate source of biometric information.
However, since some parts of the eyes may be obstructed by, e.g., eyelashes, some bits of the Iris Codes should be excluded from matching protocols.
Thus, Iris Codes are usually accompanied by a mask which removes faulty bits before matching.
In the Worldcoin infrastructure, dedicated iris capture stations (called \emph{orb}s) are used to create both the Iris Codes and the corresponding mask, while ensuring the liveness property of the user.
Matching itself is done by computing a normalized hamming distance, which is then compared to a threshold.

The main uniqueness check protocol, as given in \Cref{alg:iris_plain}, calculates whether a new Iris Code $\vec{c}\in\F_2^l$ with the mask $\vec{m}\in\F_2^l$ is similar (calculated via the hamming distance) to any Iris Code in a given database.
In the currently deployed system the Iris Code size $l$ is given as $l=12800$.
Furthermore, the deployed iris check involves additional protection against false negatives.
First, each Iris Code is rotated $r=31$ times by a small offset and each rotation is individually checked against the database.
Second, codes for both irises per person are checked against the database.
While \Cref{alg:iris_plain} does not explicitly consider these two additional checks, they can simply be modelled by executing it $2r$ times.

\begin{algorithm}[ht]
    \caption{The Iris Code Membership Protocol without MPC. It checks whether the Iris Code $\vec{c}$, under the mask $\vec{m}$ is similar to any iris in the database $C_\texttt{DB}$ under masks $M_\texttt{DB}$. $l$ is the size of the Iris Codes in bits, $s$ is the number of codes in the database. \label{alg:iris_plain}}
    \begin{algorithmic}[l]
        \Require $\vec{c}\in \F_2^l, \vec{m} \in \F_2^l, C_\texttt{DB}\in \F_2^{s \times l}, M_\texttt{DB} \in \F_2^{s \times l}$
        \Ensure \texttt{true} if $\vec{c}$ is similar to an entry in the DB, \texttt{false} otherwise.
        \For{$i$ in $0 .. s$}
        \State $\vec{m}' \gets \vec{m} \wedge M_\texttt{DB}[i]$
        \Comment Combine masks.
        \State $\texttt{ml} \gets \texttt{CountOnes}(\vec{m}')$
        
        \State $\texttt{hd} \gets \texttt{CountOnes}((\vec{c} \oplus C_\texttt{DB}[i]) \wedge \vec{m}')$
        \Comment Hamming distance.
        \If{$\texttt{hd} / \texttt{ml} < \texttt{MATCH\_RATIO}$}
        \State \Return \texttt{true}
        \Comment Iris is similar!
        \EndIf
        \EndFor
        
        \State \Return \texttt{false}
        \Comment No match found.
    \end{algorithmic}
\end{algorithm}

In the next sections, we explore different methods for translating \Cref{alg:iris_plain} to the MPC setting.
First, in \Cref{sec:mpc_p1} we focus on the case where only the Iris Codes will be protected, whereas the accompanying masks are assumed to be known in plain by all computing parties.
Then, in \Cref{sec:mpc_p2} we extend the solution to also protect the masks.
Finally, we give some benchmarks in \Cref{sec:benchmarks}.

\section{First MPC Protocol: Protecting the Iris Codes} \label{sec:mpc_p1}

In the first version of the MPC protocol, we aim to protect the Iris Codes while the corresponding masks are allowed to stay public.
Thus, both the new Iris Code $\vec{c}$ should be protected from the computing parties, as well as the Iris Codes in the database $C_\texttt{DB}$.
The masks $\vec{m}$ and $M_\texttt{DB}$ are assumed to be known by the parties.
This section also serves as a stepping stone for the version of the protocol in \Cref{sec:mpc_p2}, where masks are also protected.

\subsection{Efficient Hamming Distance Calculation} \label{sec:hd}

One of the core operations of the Iris Code Membership protocol is the calculation of the hamming distance of the two binary Iris Code vectors.
The main issue with this computation in MPC is that the hamming distance requires an XOR operation, i.e., an operation in $\F_2$, followed by counting the ones to get a result in a larger ring $\Z_t$.
Finally, after the comparison with the threshold, the result will be a boolean value again.
Calculating the hamming weight of a binary vector in MPC in a trivial fashion would thus require communication that is linear in the length of the vector $l$.
This comes from either translating the shared bits to arithmetic shares which requires communication for each bit, or from counting the ones in a binary circuit, which also requires communication at least linear in the length $l$.
However, if we have the precondition that the input vectors are already shared
over a larger ring $\Z_t$ instead of $\F_2$, we can rewrite the hamming distance calculation to

\begin{equation}
    \mathsf{hd}(\vec{a}, \vec{b}) = \sum_i a_i + \sum_i b_i - 2 \cdot \langle a_i,b_i\rangle \,. \label{eq:hd}
\end{equation}

This reduces the calculation of the hamming distance to two sums which can be computed without party interaction in secret-sharing based MPC protocols, as well as a dot-product of two vectors.
The calculation of this dot-product dominates the complexity of the hamming-distance operation, and we therefore want to use MPC protocols that support efficient dot-product evaluations.
Protocols that have a honest-majority security assumption (e.g., Shamir secret sharing \cite{DBLP:journals/cacm/Shamir79} and replicated sharing protocols such as ABY3 \cite{DBLP:conf/ccs/MohasselR18}) can support dot-products that require communication which is independent of the size of the vectors, allowing us to efficiently realize the hamming distance calculation in MPC.
This optimization alone reduces communication from $\approx 3\,$GB when using binary circuits to just $2\,$MB for $l=12800, t=2^{16}$ and a database size of $s=1\,000\,000$.

\subsubsection{Masked Bitvectors: Reducing the Workload} \label{sec:masked}

When looking at the equation $\texttt{hd} \gets \texttt{CountOnes}((\vec{c} \oplus C_\texttt{DB}[i]) \wedge \vec{m}')$, i.e., \texttt{hd} is calculated on bitvectors and masks, one can apply a further optimization by changing the representation of masked Iris Codes.

First, we introduce a new type, dubbed \emph{masked bit}, consisting of three states $\{\texttt{T}, \texttt{U}, \texttt{F}\}$, where \texttt{U} indicates that the mask bit is not set, \texttt{F} indicates a set mask bit while the code bit is not set, and \texttt{T} indicates both mask and code bits are set.
We represent this new type as  $(\texttt{T}, \texttt{U}, \texttt{F}) = (-1, 0, 1)$.
One can translate a bit $a\in\F_2$ with the mask $m_a\in\F_2$ to the masked bit representation $a' \in \{-1, 0, 1\}$ by computing $a' = m_a - 2 \cdot (a \wedge m_a)$ in a larger ring (using the canonical embedding of signed integers in such a ring).

By using this representation, one can rewrite the comparison $\texttt{hd} < \texttt{MATCH\_RATIO} \cdot \texttt{ml}$ to a version using masked bitvectors, i.e.,

\begin{equation*}
    \langle c', C'_\texttt{DB}[i]\rangle > (1 - 2 \cdot  \texttt{MATCH\_RATIO}) \cdot \texttt{ml}\,,
\end{equation*}
removing two sums from the computation, improving performance.
We refer to \Cref{ap:masked_bitvectors} for more details.
Furthermore, only the left-hand-side of the resulting comparison, i.e., the dot product, depends on shared values, the right-hand-side can be calculated from public values only.

\subsection{Efficient Comparison} \label{sec:comparison}

Since a comparison $a < b$ is equal to an MSB-extraction $\texttt{msb}(a-b)$ if the sizes of $a,b$ are chosen to not produce an overflow in the chosen signed integer representation (see \Cref{lemma:overflow}), a core building block is an MSB-extraction protocol.
This subprotocol requires to change the sharing type of additive shares over $\Z_t$ to boolean shares over $\F_2^{\lceil\log_2(t)\rceil}$.
3-party replicated sharing protocols, such as ABY3 \cite{DBLP:conf/ccs/MohasselR18}, are amongst the most efficient MPC protocols for converting between shares over bits and larger rings.

\paragraph{Accumulating the Resulting Bits.}

If one wants to prevent leaking which Iris Code in the database was similar to the query, one can accumulate all resulting comparison bits by setting them to true if at least one comparison was true.
This operation is equivalent to many boolean OR operations.
In MPC, one can compute $x \vee y = x \xor y \xor (x \wedge y)$.
To reduce the number of communication rounds, we evaluate the accumulation of all bits in a binary tree.

\subsection{MPC Protocols}

After discussing the subprotocols in the previous sections we give the final MPC algorithm in \Cref{alg:iris_shared}.
The workflow is the following: The orb station which creates a new Iris Code $\vec{c}\in\F_2^l$ with mask $\vec{m}\in\F_2^l$ secret shares each bit of $\vec{c}$ (using the masked representation of \Cref{sec:masked}) over a larger ring $\Z_t$ to the MPC nodes while sending the mask in plain.
Then, the MPC nodes evaluate \Cref{alg:iris_shared} to produce a shared result bit, which is true if the new Iris Code is similar to an existing one.
This bit is then reconstructed at a chosen output node.

\begin{algorithm}[ht]
    \caption{The Iris Code Membership Protocol with MPC. It checks, whether the secret-shared Iris Code $\shared{\vec{c}}$ (represented as masked bitvector), under the public mask $\vec{m}$ is similar to any secret-shared Iris Code in the database $\shared{C_\texttt{DB}}$ (represented as masked bitvector) under public masks $M_\texttt{DB}$. $l$ is the size of the Iris Codes in bits, $s$ is the number of codes in the database. \label{alg:iris_shared}}
    \begin{algorithmic}[l]
        \Require $\shared{\vec{c}}\in \Z_t^l, \vec{m} \in \F_2^l, \shared{C_\texttt{DB}} \in \Z_t^{s \times l}, M_\texttt{DB} \in \F_2^{s \times l}$
        \Ensure Sharing of \texttt{true} if $\vec{c}$ is similar to an entry in the DB, \texttt{false} otherwise.
        \State $\shared{\vec{b}}\gets \shared{\vec{0}} \in \F_2^s$
        \For{$i$ in $0 .. s$}
        \State $\texttt{ml} \gets \texttt{CountOnes}(\vec{m} \wedge M_\texttt{DB}[i])$
        \Comment Combine masks.
        
        \State $\shared{\texttt{hd}} \gets \langle \shared{\vec{c}},  \shared{C_\texttt{DB}[i]}\rangle$
        \Comment Hamming distance.
        \State $\shared{b[i]} \gets \texttt{MSB}((1 - 2 \cdot  \texttt{MATCH\_RATIO}) \cdot \texttt{ml} - \shared{\texttt{hd}})$
        \Comment Comparison
        \EndFor
        
        \State \Return $\texttt{OrTree}(\shared{\vec{b}})$
        \Comment Aggregate the resulting bits
    \end{algorithmic}
\end{algorithm}

\subsubsection{Baseline: Semi-honest ABY3} \label{sec:semi_aby3}

Since 3-party replicated honest majority protocols, such as ABY3 \cite{DBLP:conf/ccs/MohasselR18} provide both an efficient dot product, as well as efficient arithmetic to binary conversion for secret shares, we will use the semi-honest variant of ABY3 as a baseline for further discussions.
We refer to \Cref{ap:aby3} for a small Introduction to ABY3. Importantly, computing a dot-product requires the same amount of communication as a single multiplication and requires a CPU workload of only two plain dot-products in our use case.

\paragraph{Sharing Ring and MSB-extraction.} \label{sec:msb}

As mentioned in \Cref{sec:comparison}, we will use the efficient arithmetic-to-binary conversion of ABY3 to evaluate the comparison $\shared{b[i]} \gets \texttt{MSB}((1 - 2 \cdot  \texttt{MATCH\_RATIO}) \cdot \texttt{ml} - \shared{\texttt{hd}})$ in \Cref{alg:iris_shared}.
Consequently, we need to ensure that we choose the bitsize $k$ of the ring $\Z_{2^k}$ used for sharing to be large enough, such that the subtraction $(1 - 2 \cdot  \texttt{MATCH\_RATIO}) \cdot \texttt{ml} - \shared{\texttt{hd}}$ does not produce overflows in the canonical representation of signed integers in the ring.

\begin{lemma} \label{lemma:overflow}
    The comparison of $\texttt{hd} > f \cdot \texttt{ml}$ for a real number $f\in[0,1]$, \texttt{hd} being the dot product of two masked bitvectors of size $l$, and \texttt{ml} being the dot product of two bitvectors of size $l$, is equivalent to $\texttt{MSB}(\lceil f \cdot \texttt{ml} \rceil - \texttt{hd})$  when calculated over $\Z_t$, if $l<\frac{t}{4}$ and $l<t-2^{\lceil\log_2(t)\rceil - 1}$.
\end{lemma}

For a proof of \Cref{lemma:overflow} we refer to \Cref{ap:proof_lemma_overflow}. Using \Cref{lemma:overflow}, we choose $\Z_t$ with $t=2^{16}$ as the sharing ring, since both conditions are fulfilled for $l=12800$.

To extract the MSB of a share $\shared{x}\in\Z_{2^{16}}$, we transform it to boolean shares over $\F_2^{16}$.
In ABY3 this is done by interpreting the share $\shared{x}=(x_1, x_2, x_3)$ (where $x = x_1 + x_2 + x_3\mod 2^{16}$) as three trivial boolean shares and adding them via a 16-bit binary adder circuit in MPC, which implicitly reduces the result mod $2^{16}$.
We give the algorithmic description of the MSB-extraction procedure in \Cref{ap:algorithms}.
When using a ripple-carry adder (\Cref{alg:BinAdd_RCA}), which requires the least amount of AND gates, MSB extraction requires 29 AND gates in 15 communication rounds.

\subsubsection{Shamir Sharing: Reducing the database and dot products}

As discussed in \Cref{ap:aby3}, an ABY3 share consists of two additive shares.
Consequently, since one has to share each bit of the Iris Code as a sharing over $\Z_{2^{16}}$, the database increases from $s\cdot l$ bits to $2\cdot s \cdot 16 \cdot l = 32 \cdot s\cdot l$ bits.  Furthermore, for each MPC dot product, two plain dot products have to be calculated, a particularly undesirable property since the dot products will dominate the runtime on a CPU for large databases.

Luckily, one can counteract the disadvantages of doubled share size and computation cost through switching to Shamir secret sharing \cite{DBLP:journals/cacm/Shamir79}.
In Shamir sharing, secrets are shared as points on random polynomials of degree $d$, where the secret itself is located in the constant term of the polynomial.
The secrets can be reconstructed from the shares by, e.g., Lagrange interpolation of $d+1$ shares, which is a linear operation.
Similar to additive sharing, linear operations can be performed on the shares without requiring party interaction and without changing the degree $d$ of the underlying sharing polynomial.
While multiplications can be implemented as direct multiplications of the shares, they increase the degree of the underlying polynomial to $2\cdot d$, i.e., about twice as many parties are needed for reconstruction.
Consequently, multiplications are usually followed by a degree reduction step, which requires party interaction.
However, similar to ABY3, one does not immediately have to reduce the degree after each multiplication in a dot-product evaluation.
Consequently, by computing additions locally prior to degree reduction, a dot product over Shamir shares can be implemented with communication equivalent to that of a single multiplication.

\paragraph{Shamir Sharing over Rings.}

One difference between using ABY3 and Shamir sharing, however, is that one cannot instantiate Shamir secret sharing over $\Z_{2^{16}}$ directly --- instead, one usually has to instantiate it over some prime field $\F_p$.
This is undesirable in our case because the MSB-extraction becomes more costly in a prime field since the mod $p$ reduction has to be embedded into the binary addition circuit evaluated during the arithmetic-to-binary-conversion.

The reason why Shamir cannot trivially be applied for polynomials over general rings is the following.
For interpolating the constant term $p[0]$ of the degree $d$ polynomial $p[X]$ from $d+1$ points $\{p[x_1], p[x_2], \dots p[x_{d+1}]\}$, one calculates $p[0] = \sum_{i=1}^{d+1} \lambda_{x_i} \cdot p[x_i]$, where $\lambda_{x_i} = \prod_{j=1,j\neq i}^{d+1} \frac{x_j}{x_j -x_i}$.
In particular, the pairwise differences $x_j - x_i$ of the evaluation points  need to be invertible in order to instantiate Shamir secret sharing in this way.
Furthermore, an ``exceptional sequence'' of evaluation points satisfying this property needs to have length $n+1$ to support n-party Shamir secret sharing; $n$ for creating the shares for the parties and one additional point to embed the secret (usually 0).
Evidently, in power-of-two-rings $\Z_{2^k}$, the scheme can't be realized because only odd numbers are invertible, so an exceptional sequence can have at most two elements -- insufficient for Shamir secret sharing even for two parties.

To increase the possible size of an exceptional sequence of evaluation points, one can instantiate Shamir secret sharing over so-called \emph{Galois rings}: rings of the form\footnote{For the fully general definition of a Galois ring, 2 should be replaced with a prime number $p$.} $\Z_{2^k}[X]/Q[X]$ where $Q[X]$ is a polynomial whose projection (by coefficients) into $\Z_2[X]$ is irreducible.
In other words, coefficients, shares, and evaluation points of Shamir secret sharing are now elements in a polynomial quotient ring whose coefficients are in $\Z_{2^k}$.
While this has the effect of increasing the number of exceptional points to $2^d$ where $d$ is the degree of $Q[X]$ (called the \emph{extension degree} of the Galois ring), it also leads to a computational performance penalty due to replacing ring arithmetic in $\Z_{2^k}$ with polynomial arithmetic.

\paragraph{Optimized 3-party Shamir over Galois Rings.}

For our specific protocol, we use Galois ring Shamir sharing only to compute a dot product, while the remainder of the protocol (i.e., the MSB extraction) makes use of replicated sharing.
This setting allows the following performance optimization.
First, we instantiate a Galois ring $\Z_{2^{16}}[X]/(X^2-X-1)$, which has an exceptional sequence of size $2^d=4$, enough for 3-party Shamir secret sharing.
Next, we pack two consecutive iris bits into one Galois ring element $g_i = c_{2i} + c_{2i+1} X$.
The product of two Galois ring elements $g_i$ and $h_i$ whose coefficients encode values in $\Z_{2^{16}}$ is given by
\begin{align}\label{eq:galois_mul}
    g_i \cdot h_i & = (c_{2i} + c_{2i+1} X) \cdot (d_{2i} + d_{2i+1} X) \mod (X^2-X-1) \notag                          \\
                  & = (c_{2i} d_{2_i} + c_{2i+1}d_{2i+1}) +  (c_{2i} d_{2_i + 1} + c_{2i+1}d_{2i} +c_{2i+1}d_{2i+1}) X
\end{align}
Observe in particular that the constant term of the result of \Cref{eq:galois_mul} is the dot product of the packed inputs.

Furthermore, when performing this multiplication on Shamir shares, the result is a share on a polynomial of degree $2d$ where $d$ is the degree of the sharing polynomial of the inputs $g_i$ and $h_i$.
This share can be transformed into a ($2d+1$)-party additive secret sharing by multiplying the share with the corresponding Lagrange coefficient $\lambda$.
Consequently, letting $[g_i]_\text{Shamir}$ denote a degree-$d$ Shamir sharing of $g_i\in\Z_{2^{16}}[X]/(X^2-X-1)$, and $[c_i]_\text{Add}$ a ($2d+1$)-party additive sharing of $c_i\in\Z_{2^{16}}$, we can embed the Shamir-to-additive translation into the multiplication:
\begin{align}
    (\lambda \cdot [g_i]_\text{Shamir}) & \cdot [h_i]_\text{Shamir} =  (\lambda \cdot [c_{2i} + c_{2i+1} X]_\text{Shamir}) \cdot [d_{2i} + d_{2i+1} X]_\text{Shamir} \notag \\
                                        & = [c_{2i} d_{2_i} + c_{2i+1}d_{2i+1}]_\text{Add} + [c_{2i} d_{2_i + 1} + c_{2i+1}d_{2i} +c_{2i+1}d_{2i+1}]_\text{Add} X
\end{align}

As a result, when precomputing $[g'_i] = \lambda\cdot [g_i]$, the constant term of $[g'_i] \cdot [h_i]$ is a ($2d+1$)-party additive sharing of the dot-product of the packed inputs. Thus, following this precomputation, one does not have to compute the full Galois ring multiplication when computing a dot product --- only the constant term is required. This optimization allows us to instantiate the protocol with Shamir secret sharing over Galois rings, while computing the dot product over $\Z_{2^{16}}$ at no extra cost beyond the precomputation for the shares $[g'_i]$.

\paragraph{Iris Membership using Shamir Sharing over Galois Rings.}

When using Shamir secret sharing instead of ABY3 the workflow changes as follows:
First, the orb shares a new Iris Code with the MPC parties by sharing each two consecutive bits as ($c_{2i} + c_{2i+1}X$) using Shamir sharing over $\Z_{2^{16}}[X]/(X^2-X-1)$, with the sharing polynomial having degree one.
After receiving the shares of the query, the MPC parties multiply each share with their Lagrange coefficient over $\Z_{2^{16}}[X]/(X^2-X-1)$.
Then, the parties can continue computing the dot product of the query with each Iris Code in the database as if all bits of the Iris Codes were shared over $\Z_{2^{16}}$.
The resulting dot products are valid 3-party additive shares which can be transformed to replicated sharing by sending the share of party $i$ to the next party $i+1$ (including re-randomization).
The remainder of the protocol, i.e., the MSB-extraction and potentially the OR-tree, are equivalent to the version using only ABY3.

As a result, using Galois Sharing reduces the database size of the server from $32\cdot s \cdot l$ using ABY3 to $16\cdot s \cdot l$, while also reducing the computational workload from two plain dot products to just one.

\begin{remark}
    In this section we present optimized dot products in Galois rings for 3-party Shamir secret sharing. In general, it is possible to extend the same techniques to support additional parties (i.e., increasing the extension degree of the Galois ring), while keeping the same efficiency gains. We leave it to future work to formalize the trick in a more generic way.
\end{remark}

\section{Second MPC Protocol: Protecting the Codes and Masks} \label{sec:mpc_p2}

The goal of the second MPC protocol is to extend \Cref{alg:iris_shared} to also protect the masks $\vec{m}\in \F_2^l$ and $M_\texttt{DB} \in \F_2^{s\times l}$.
This triggers two changes:
First, the mask bits also need to be shared over a larger ring $\Z_t$ and \texttt{ml} can be calculated by a dot product.
Second, since \texttt{ml} is unknown to the computing parties, they cannot trivially multiply it with a real value $f$ and round the result.
The second change is the reason why we need to approximate $(1 - 2 \cdot  \texttt{MATCH\_RATIO}) \approx \frac{a}{b}$ with $a\le b$ and $a,b\in\Z_{t'}$.
Then, while ensuring that the operations do not overflow, $(1 - 2 \cdot  \texttt{MATCH\_RATIO}) \cdot \texttt{ml} < \texttt{hd}$ can be calculated as $\texttt{MSB}(a \cdot \shared{\texttt{ml}} - b\cdot \shared{\texttt{hd}})$.
We depict the resulting algorithm in \Cref{alg:shared}.

\begin{algorithm}[ht]
    \caption{The Iris Code Membership Protocol with MPC. It checks, whether the secret-shared Iris Code $\shared{\vec{c}}$ (represented as masked bitvector), under the shared mask $\shared{\vec{m}}$ is similar to any secret-shared iris in the database $\shared{C_\texttt{DB}}$ (represented as masked bitvector) under shared masks $\shared{M_\texttt{DB}}$. $l$ is the size of the Iris Codes in bits, $s$ is the number of codes in the database. \label{alg:shared}}
    \begin{algorithmic}[l]
        \Require $\shared{\vec{c}}\in \Z_{t_1}^l, \shared{\vec{m}} \in \Z_{t_2}^l, \shared{C_\texttt{DB}} \in \Z_{t_1}^{s \times l}, \shared{M_\texttt{DB}} \in \Z_{t_2}^{s \times l}$
        \Ensure Sharing of \texttt{true} if $\vec{c}$ is similar to an entry in the DB, \texttt{false} otherwise.
        \State $\shared{\vec{b}}\gets \shared{\vec{0}} \in \F_2^s$
        \For{$i$ in $0 .. s$}
        \State $\shared{\texttt{ml}} \gets \langle \shared{\vec{m}},  \shared{M_\texttt{DB}[i]}\rangle$
        \Comment Combine masks.
        
        \State $\shared{\texttt{hd}} \gets \langle \shared{\vec{c}},  \shared{C_\texttt{DB}[i]}\rangle$
        \Comment Hamming distance.
        
        \State $\shared{\texttt{ml}'} \gets \texttt{lift\_ml}(\shared{\texttt{ml}})$
        \Comment Some lifting to a larger ring to prevent overflows
        
        \State $\shared{\texttt{hd}'} \gets \texttt{lift\_hd}(\shared{\texttt{hd}})$
        \Comment Some lifting to a larger ring to prevent overflows
        
        \State $\shared{b[i]} \gets \texttt{MSB}(a \cdot \shared{\texttt{ml}'} - b\cdot \shared{\texttt{hd}'})$
        \Comment Comparison
        \EndFor
        
        \State \Return $\texttt{OrTree}(\shared{\vec{b}})$
        \Comment Aggregate the resulting bits
    \end{algorithmic}
\end{algorithm}

Due to changing the comparison operation, we have to adapt \Cref{lemma:overflow} to \Cref{lemma:overflow_2} for correct bounds, such that the MSB extraction is equivalent to the comparison. The proof can be found in \Cref{ap:proof_lemma_overflow2}.

\begin{lemma} \label{lemma:overflow_2}
    The comparison of $\texttt{hd} > \frac{a}{b} \cdot \texttt{ml}$ for $a, b\in\Z_{t'}$, $a\leq b$, \texttt{hd} being the dot product of two masked bitvectors of size $l$, and \texttt{ml} being the dot product of two bitvectors of size $l$, is equivalent to $\texttt{MSB}(a \cdot \texttt{ml} - b \cdot \texttt{hd})$  when calculated over $\Z_t$, if $b \cdot l<\frac{t}{4}$ and $b\cdot l<t-2^{\lceil\log_2(t)\rceil - 1}$.
\end{lemma}

\subsection{Lifting vs. Larger Dot-Product}\label{sec:lifting_vs_large_dot}

\Cref{alg:shared} immediately shows the potential for trading off the cost of the dot product with the MSB extraction algorithm.
First, if we compute the dot products over a smaller ring, e.g., a 16-bit modulus $t$, we have to lift (in MPC) the results to a larger space (e.g., 32-bit to allow 16-bits of precision for $a$ and $b$) such that the multiplications with $a,b$ do not overflow and the MSB extraction can be used for the comparison.
Second, if we directly compute the dot products in a larger ring, we do not have to compute the lifting in MPC at the cost of more expensive dot product computations.

Thus, we will start with proposing some efficient lifting algorithms.
First, observe that lifting a multiplication is free if we change the resulting ring accordingly.
Let $[x]_t$ be an additive share mod $t$, then the multiplication (without modular reduction) $s\cdot[x]_t = [s\cdot x]_{st}$ becomes a valid sharing of $s\cdot x \mod st$ (see \Cref{alg:constlift}).
Thus, when approximating $(1 - 2 \cdot  \texttt{MATCH\_RATIO}) \approx \frac{a}{b}$ with $b=2^m$ being a power-of-two, lifting $\texttt{hd}$ to the ring $\Z_{2^{k + m}}$ can be done without interaction.
Consequently, $\texttt{hd}$ can always be computed in the smaller ring without performance loss.

\paragraph{Lifting in Power-of-Two-Rings.}

Since $a$ is different from $b=2^{m}$, we have to use a different method for lifting $[\texttt{ml}]_{2^{n}}$ to $[\texttt{ml}]_{2^{m+n}}$, which will require communication.
Our approach involves the following steps.
First, we directly interpret the sharing of \texttt{ml} as shares over $\Z_{2^{m+n}}$.
This result represents $\texttt{ml}_1 + \texttt{ml}_2 + \texttt{ml}_3$ without reduction modulo $2^{n}$, which is why we need to perform it manually.
We do this by extracting the $n$-th and $(n+1)$-th bit (when interpreting the LSB as the 0-bit) of this value, transforming them into a sharing over $\Z_{2^{n+m}}$ using a bit-injection algorithm to get $[b_{n}]_{2^{m+n}}$ and $[b_{n+1}]_{2^{n+m}}$ and calculating $[\texttt{ml}]_{2^{m+n}} = [\texttt{ml}']_{2^{n+m}} - 2^{n} \cdot [b_{n}]_{2^{m+n}} - 2^{n+1} \cdot [b_{n+1}]_{2^{m+n}}$.
Furthermore, one can optimize the bit-injection step to inject into a smaller ring and use \Cref{alg:constlift} to perform the multiplication with $2^{n}$ to save some bits of communications.
The resulting algorithm can be found in \Cref{alg:Lift} in \Cref{ap:algorithms}.

Using \Cref{lemma:overflow_2}, we can set the ring $\Z_t$ in which we calculate the comparison as MSB extraction.
Since $l=12800$, and $b=2^m$ for the efficient lifting of \texttt{hd} (where $m$ is chosen for enough accuracy of the approximation of the threshold via $\frac{a}{b}$), we can simply choose $t=2^{16+m}$.

\subsection{Final MSB Extraction}

Similar to \Cref{sec:mpc_p1}, we can use \Cref{alg:BitExtract} (in \Cref{ap:algorithms}) to extract the MSB.
Consequently, for $k$-bit moduli, this step requires $2k-3$ bits of communication in $k-1$ rounds.

\section{Benchmarks} \label{sec:benchmarks}

In this section we give some benchmarks for our proposed protocols.\footnote{The full implementation is available at \url{https://github.com/worldcoin/iris-mpc}.}
While we give full end-to-end benchmarks of a deployed system, we also focus on comparing different instantiations of the two main parts of the protocol, i.e., the dot product evaluation as well as the comparisons via the MSB-extraction in different settings.
On a high-level, these two subprotocols have completely different performance characteristics.
First, the dot products have larger memory requirements, are more expensive on the CPU, and do not require network interaction except for the resharing in one communication round in the end.
On the other hand, the comparisons (including lifting in MPC), have more communication rounds with comparably small CPU requirements in between.

In the next sections, we focus on 16-bit precisions of the $a/b$ approximation of the threshold. Thus, the ring for comparison will extend from 16-bit to 32.

\subsection{Dot-Products}

The evaluation of the dot-products in MPC is the computationally most expensive part of the protocol. After this first phase, the input vectors of length 12800 get reduced to a single value, which is then used in the comparison phase. We implemented CPU and GPU variants of the dot product and benchmarked the throughput of these variants on various large-scale instances that can be rented from AWS. We evaluate the CPU implementation on Graviton3 instances, and the GPU implementation on both NVIDIA A100s and H100s.

\paragraph{CPU Implementation.}

The Neoverse1 cores used in Graviton3 instances are based on the ARMv8.4 architecture and support the 16-bit SVE \texttt{UDOT} instruction, which can be used to efficiently compute the dot product in 16-bit rings.

For larger rings, no dedicated dot-product instructions exist and an implementation of the dot-product utilizing both SVE and scalar instructions is used instead. We give the performance of the CPU dot product in \Cref{tab:cpu_gpu_dot_perf}. Note that these numbers only concern the inner loop and do not include effects of the cache hierarchy. We therefore expect that the performance of a full production solution will be slightly lower.

\paragraph{GPU Implementation.}

Even though our proposed protocol for matching Iris Codes has nothing to do with deep learning, we can still leverage the capabilities of modern GPUs to compute the pairwise hamming distances of Iris Codes through large matrix multiplications, the prime use case for GPU acceleration. Unfortunately, the General Matrix Multiplications (GEMM) algorithms in cuBLAS are not directly applicable to our problem, since the only configuration with suitable precision (at least for the 16 bit setting) would be FP64\footnote{\url{https://docs.nvidia.com/cuda/cublas/\#cublasgemmex}}. The only integer data type supported in cuBLAS is \texttt{int8} for the operands and \texttt{int32} for accumulation of the result. However, by applying decomposition of the operands into smaller data types \cite{DBLP:conf/sp/TanKTW21, DBLP:conf/uss/WatsonWP22}, we are able to use the \texttt{int8} Tensor operations (taking care of any deviations due to the use of signed integers with custom CUDA kernels).
Notably, since we operate in a 16 bit or 32 bit unsigned integer ring, one only needs 3 or 10 individual \texttt{int8 GEMM} operations respectively, since the higher order cross-terms cancel out in the modular reduction.
The benchmarks were conducted using \texttt{cuBLAS} and the Rust library \texttt{cudarc}\footnote{\url{https://github.com/coreylowman/cudarc}}, and similar to the CPU version, only contains the matrix multiplication and no other parts of the protocol. The results are depicted in \Cref{tab:cpu_gpu_dot_perf}.

\begin{table}[th]
    \caption{Performance of dot-product implementations on a 64-core Graviton3 instance vs. Nvidia A100 and H100 GPUs, measured in millions of length 12800 dot-products/s.}
    \label{tab:cpu_gpu_dot_perf}
    \centering
    \begin{tabular}{lrrrr}
        \toprule
                  & \multicolumn{2}{c}{Replicated Sharing} & \multicolumn{2}{c}{Galois Shamir}                                 \\
                  & $\Z_{2^{16}}$                          & $\Z_{2^{32}}$                     & $\Z_{2^{16}}$ & $\Z_{2^{32}}$ \\
        \midrule
        Graviton3 & 49                                     & 24                                & 98            & 48            \\
        \midrule
        A100      & 1364                                   & 496                               & 2728          & 992           \\
        H100      & 3146                                   & 1162                              & 6292          & 2324          \\
        \bottomrule
    \end{tabular}
\end{table}

\begin{figure}
    \caption{Cost-normalized throughput of the different dot-product implementations.}
    \label{fig:cost_normalized_throughput}
    \centering
    \includegraphics[width=0.8\textwidth]{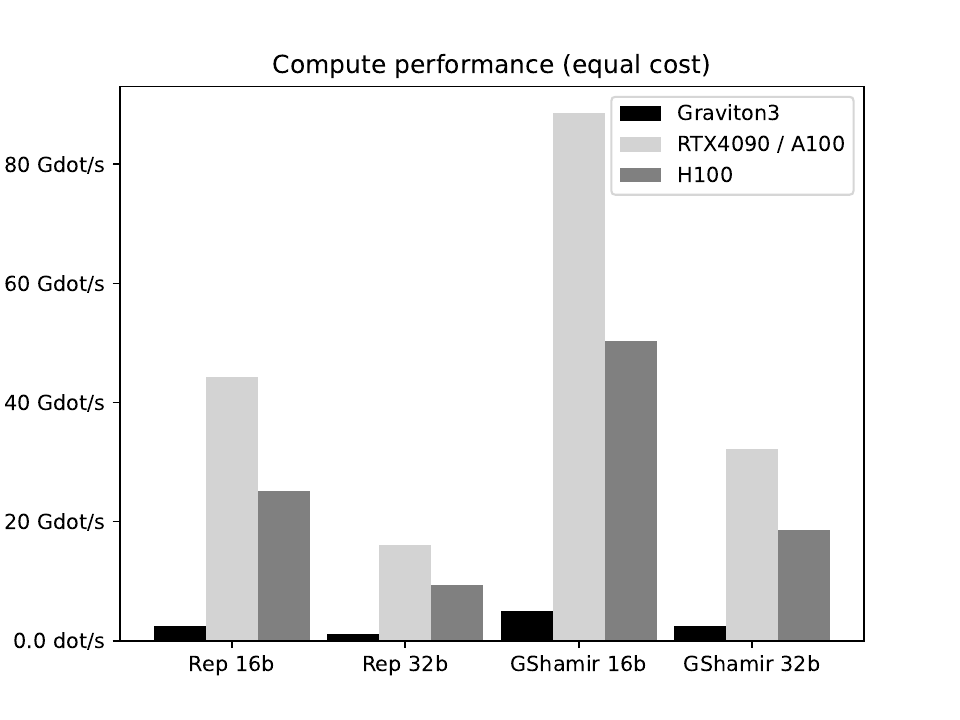}
\end{figure}

Comparing the performance numbers in \Cref{tab:cpu_gpu_dot_perf} we see that GPUs excel at workloads like dot-products and vastly outperform even CPUs with dedicated dot-product instructions.
However, the main metric that is of interest for deployment is the cost-normalized throughput of the instances, and so we compare the 3 different implementations in \Cref{fig:cost_normalized_throughput}, based on the monthly upfront cost of \texttt{hpc7g.16xlarge}, \texttt{p4d.24xlarge} and \texttt{p5.48xlarge} instances on AWS.
Still, the GPU implementation is the most cost-effective solution for the evaluation of the dot-products.

\subsection{Comparisons via MSB-Extraction} \label{sec:msb_bench}

The second phase of the protocol differs significantly from the dot-product evaluation, since the MPC protocols for the threshold comparison require many communication rounds.
We benchmark and compare four different settings:
First, we distinguish between knowing the masks in \textit{plain} (\Cref{sec:mpc_p1}) and the protocols from \Cref{sec:mpc_p2}.
For the latter, we compare the cases where: (i) both $\texttt{hd}$ and $\texttt{ml}$ are computed over 16-bit and we lift both values to a 32-bit sharing (using $b=2^{16}$, dubbed \textit{MPCLift}), (ii) we compute $\texttt{ml}$ over 32-bit such that we only have to lift \texttt{hd} with $b=2^{16}$, which can be done for free (\textit{ConstLift}), (iii) we compute both  $\texttt{ml}$ and $\texttt{hd}$ over 32-bit (\textit{NoLift}) and use 16-bit $a,b$ approximations.
Furthermore, we give the benchmarks for accumulating all results into one shared output bit via an Or-tree evaluation.
The benchmarks are given in \Cref{tab::bench_msb} for a batch sizes of 100k $\texttt{hd} < \texttt{ml} \cdot \texttt{MATCH\_RATIO}$ evaluations in parallel.

All benchmarks are obtained on a machine with an AMD Ryzen 9 7950X CPU (4.5 GHz), where each party is executed on an individual thread and the parties are connected via a network connection on the same local host. While a real-world deployment would imply a different networking setup, the performance requirements imply that any cluster of MPC nodes would likely be connected via multiple high-speed networking links in the same data-center, which should not significantly impact the performance of the protocol.

\begin{table}[ht]
    \centering
    \caption{Comparison of the different protocols for doing the $\texttt{hd} < \texttt{ml} \cdot \texttt{MATCH\_RATIO}$ comparison 100k times in parallel in MPC.
        The parties are run on a single thread each on the same CPU, with localhost networking.
        Data gives the communication as amount of kB send per party.
        Results are averaged over 1000 executions.}
    \label{tab::bench_msb}
    \small
    \begin{tabular}{lrrr} \toprule
        \multicolumn{1}{c}{Protocol} & \multicolumn{1}{c}{Runtime} & \multicolumn{1}{c}{Throughput} & \multicolumn{1}{c}{Comm} \\
                                     & \multicolumn{1}{c}{$ms$}    & \multicolumn{1}{c}{M\,el/$s$}  & \multicolumn{1}{c}{kB}   \\
        \midrule
        Plain Mask                   & 3.18                        & 31.4                           & 362                      \\
        \midrule
        MPCLift                      & 14.47                       & 6.9                            & 2\,138                   \\
        ConstLift                    & 5.32                        & 18.8                           & 763                      \\
        NoLift                       & 5.27                        & 19.0                           & 763                      \\
        \midrule
        Or Tree                      & 0.87                        & 114.9                          & 12                       \\
        \bottomrule
    \end{tabular}
\end{table}

\subsection{Discussion}

As already discussed in \Cref{sec:lifting_vs_large_dot}, the protocol allows for a tradeoff between the work done in the dot-product phase and the comparison phase.
Looking at CPU dot-product performance in \Cref{tab:cpu_gpu_dot_perf} first, we can see that 16-bit rings are much more efficient than 32-bit ones.
Furthermore, using Shamir sharing over Galois rings for this step offers additional performance benefits, since the multiplication boils down to a single multiplication, whereas we need to perform a second dot-product when using the replicated sharing. The ladder doubles the cost for replicated sharing, while also introducing additional memory pressure.

On the other hand, the MPC lifting step required for the comparison phase adds additional overhead, as the throughput of the comparison phase is nearly 3x as large when values are already in the larger ring (see \Cref{tab::bench_msb}).
Finally, the Or-tree evaluation is a very lightweight operation with negligible overhead compared to the rest of the protocol.

When considering the GPU numbers, we can observe that the MSB-extraction phase is the bottleneck of the protocol, as a single full 64-core Graviton instance is not sufficient to keep up with the throughput of the GPU-accelerated dot-product phase for the 16-bit ring (observe that \Cref{tab::bench_msb} reports single-threaded performance).

Regarding the communication overhead, we see that the strategy using MPC lifting requires to send 21 bytes of communication per comparison. As a single core can produce a throughput of 7 million comparisons, this amounts to 147 MB/s bandwidth requirement per core.
For a full-scale production deployment managing to process 10 queries per second against a database of 10 million Iris Codes (considering 2 irises per user and also 31 rotations per iris query) the production cluster would require $10M \cdot 10 \cdot 2 \cdot 31 \cdot 21$ bytes/s = 130 GB/s of total bandwidth, further highlighting the scale of the problem.

We choose to use 16-bit Galois rings for deployment which offer the highest throughput and use the \textit{MPCLift} version of the MSB extraction to perform the comparison in a 32-bit ring. We give the results for implementing this setting on GPUs in the next section

\subsection{GPU Implementation for Deployment}

To deploy the final protocol we implement the full Iris matching protocol on GPUs using cudarc.
While the dot products are very well suited for GPU implementation, the lifting and MPC extraction subprotocols have many communication rounds with comparably low computational effort in between.
To avoid high performance penalties from moving data between CPUs and GPUs, we use NCCL\footnote{\url{https://developer.nvidia.com/nccl}} to let GPUs directly access network interfaces. Consequently, all computations and communications are directly handled by GPUs and we essentially instantiate a whole MPC computation where the computation/communication on all parties are performed on graphic cards without ever moving data from/to CPUs.

For deployment, we choose to instantiate each MPC party using AWS P5. Thus, each party has access to 8 H100 GPUs and the parties are connected by network connections with (theoretically) up to 3.2 Tbps.

In this implementation we take the real world considerations from \Cref{sec:plain} into account, i.e., a query consists of irises derived from both eyes of a person, and both irises are rotated 31 times. Thus, a single query consists of 62 Iris Code comparisons. Additionally, we allow querying multiple persons in batches. Concretely, for the benchmark (see \Cref{tab:bench_gpu}) we consider a batch size of 32 persons, which corresponds to comparing a batch of 1984 Iris Codes to the database. Furthermore, the 64 eyes (2 eyes of 32 persons) in the batch are, in addition to matching with the database, also matched with each other to ensure uniqueness in the batch as well. Our benchmarks also include this inner-batch-matching.
The database size in this benchmark is $4\,194\,304$.

\begin{table}[ht]
    \centering
    \caption{Performance of the deployed GPU implementation of our protocols for matching a batch of 1984 Iris Codes against a database of $4\,194\,304$ Iris Codes.}
    \label{tab:bench_gpu}
    \small
    \begin{tabular}{lrrr} \toprule
        \multicolumn{1}{c}{Protocol}       & \multicolumn{1}{c}{Runtime} & \multicolumn{1}{c}{Throughput} \\
                                           & \multicolumn{1}{c}{$ms$}    & \multicolumn{1}{c}{G\,el/$s$}  \\
        \midrule
        All Dot-Products (incl. Resharing) & 370.57                      & 22.46                          \\
        Threshold Comparisons              & 1569.23                     & 5.30                           \\
        \midrule
        Total                              & 1939.80                     & 4.29                           \\
        \bottomrule
    \end{tabular}
\end{table}

\Cref{tab:bench_gpu} shows a massive gain in throughput achieved by massive parallelization from GPU clusters. While our single-threaded CPU implementation achieves a throughput of 690 thousand Iris Code comparisons per second, the GPU version using three AWS P5 instances achieves 4.29 billion comparisons per second, a performance gain of about 4 orders of magnitude.
Furthermore, these benchmark confirm again that the threshold comparison is the bottleneck in this implementation, achieving 4x less throughput than the dot products.
Nonetheless, the benchmark shows that the implementation is fast enough for the performance requirements for real world deployment by Worldcoin.

\section{Conclusion}

In this work we have shown that by utilizing efficient dot-products in \emph{honest-majority} MPC protocols we can improve the performance of secure Iris Code matching significantly. The resulting party-local dot-product workload is also a prime candidate for GPU acceleration, leading to even further speedups.
We utilize ideas from ABY3 \cite{DBLP:conf/ccs/MohasselR18} and similar protocols to switch representations of the data between different (Galois) rings and binary fields to optimize the performance of the threshold comparison compared to naive approaches and thus significantly outperform previous works.

Our benchmarks show, that when using only a single CPU core for the setting of Shamir sharing over a Galois ring $\Z_{2^{16}}[X]/(X^2-X-1)$, the dot-product implementation followed by the \texttt{MPCLift} comparison strategy allows us to compare a query Iris Code against over 690 thousand entries in the database per second.
In comparison to Janus, their SHE-Janus solution using somewhat homomorphic encryption manages to compare two query Iris Codes against a database of 8\,000 Iris Codes in 40 seconds, which is a throughput of 400 comparisons per second.
This is a factor of 1\,725 less throughput than our solution, even without considering GPU acceleration.
Furthermore, the used Iris Codes are only of length 2048, while we consider Iris Codes of length 12800, allowing for much better accuracy in the matching process.
Comparing to the SMC-Janus solution, which is based on garbled circuits, the performance is even worse, with a throughput of about 50 comparisons per second and a communication of 1\,GB for comparing two Iris Codes against a database of 1\,000 Iris Codes (where our protocol would only communicate a few tens of kilobytes).
The main reason for the performance difference to Janus is the fact that our protocol can utilize both the computational efficiency of the dot-product operation as well as its communication efficiency in honest-majority MPC protocols, which is not possible when using garbled circuits. As a consequence, MPC-based solutions are more competitive than solutions based on (somewhat) homomorphic encryption, since both solutions have communication which is independent to the size of the iris codes, while SHE traditionally introduces significantly larger overhead on CPUs.

Finally, to improve performance for a real world deployment even further, we implement the full MPC protocol on GPUs and achieve an astounding throughput of 4.29 billion Iris Code comparisons per second.
This solution will be deployed in Worldcoins World ID infrastructure, providing privacy for Iris Codes derived from millions of signed up users.

\subsection*{Acknowledgment}

This work was partially supported by the Human Collective Grant Program -- Wave 0 by the Worldcoin Foundation.\footnote{\url{https://worldcoin.org/community-grants}}

\bibliographystyle{alpha}
\bibliography{dblp,bib}

\newpage
\appendix
\section{Missing Proofs} \label{ap:proof}

In this section we give the proofs omitted from the main body of the paper.

\subsection{Proof of \Cref{lemma:overflow}} \label{ap:proof_lemma_overflow}

\begin{proof}
    \texttt{ml} is the dot product of two $l$-bit vectors, whereas \texttt{hd} is calculated as the dot product of two masked bitvectors of size $l$.
    Consequently, we can bound the ranges $\texttt{ml} \in [0, l]$ and $\texttt{hd} \in [-l, l]$.
    Furthermore, since $f$ is a real number between 0 and 1, $f \cdot \texttt{ml}$ has the same bounds as \texttt{ml}.
    Consequently $\lceil f\cdot \texttt{ml} \rceil - \texttt{hd}\in [-l, 2l]$.
    Now, we need to ensure that $\Z_t$ is big enough to represent this ranges in signed integers, i.e., $l < \frac{t}{2}$ for the negative range as well as $2l < \frac{t}{2}$ for the positive one.
    Together, we have $l < \frac{t}{4}$.
    Furthermore, we need to ensure that a positive value $x\in[0, 2l]$ leads to a MSB of 0, which is the case if $2l < 2^{\lceil\log_2(t)\rceil - 1}$ which is already ensured if $l < \frac{t}{4}$.
    Finally, negative values $x\in[-l,0)$ need to have the MSB set, which is the case if $l<t-2^{\lceil\log_2(t)\rceil - 1}$.
\end{proof}

\subsection{Proof of \Cref{lemma:overflow_2}} \label{ap:proof_lemma_overflow2}

\begin{proof}
    \texttt{ml} is the dot product of two $l$-bit vectors, whereas \texttt{hd} is calculated as the dot product of two masked bitvectors of size $l$.
    Consequently, we can bound the ranges $\texttt{ml} \in [0, l]$ and $\texttt{hd} \in [-l, l]$.
    Consequently, $a\cdot \texttt{ml}$ is in the range $[0, a\cdot l]$ and $b\cdot \texttt{ml}$ is in the range $[-b\cdot l, b\cdot l]$.
    Thus, $a\cdot \texttt{ml} -  b\cdot \texttt{hd}\in [-b\cdot l, (a+b)\cdot l]$.
    Now, we need to ensure that $\Z_t$ is big enough to represent this ranges in signed integers, i.e., $b\cdot l < \frac{t}{2}$ for the negative range as well as $(a+b)\cdot l < \frac{t}{2}$ for the positive one.
    Since $a\cdot l \geq 0$, the latter statement implies the first one.
    Furthermore, we need to ensure that a positive value $x\in[0, (a+b)\cdot l]$ leads to a MSB of 0, which is the case if $(a+b)\cdot l < 2^{\lceil\log_2(t)\rceil - 1}$ which is already ensured if $(a+b)\cdot l < \frac{t}{2}$.
    Since $a\leq b$ we have $(a+b)\cdot l \leq 2\cdot b\cdot l$, thus $b\cdot l < \frac{t}{4}$ implies $(a+b)\cdot l < \frac{t}{2}$.
    Finally, negative values $x\in[-b\cdot l,0)$ need to have the MSB set, which is the case if $b\cdot l<t-2^{\lceil\log_2(t)\rceil - 1}$.
\end{proof}

\section{Details on Masked Bitvectors} \label{ap:masked_bitvectors}

We extend logical operations, such as AND and XOR operations, such that they produce \texttt{U} at the output if at least one input is \texttt{U}.
Consequently, an XOR operation between two bits that considers their masks is equal to a multiplication:
Let $a,b\in\F_2$ be bits, $m_a,m_b\in\F_2$ be their masks, and $a',b' \in \{-1, 0, 1\}$ be their masked representation, then $a' \cdot b'$ corresponds to $(a\oplus b) \wedge m_a \wedge m_b$.
In the original bit-representation, the \texttt{CountOnes} operation counts how many bits are set.
In the masked bit representation, this operation corresponds to counting how many elements are in the state \texttt{T}. Observe that $a'^2 = a' \cdot a'$ results in $1$ if the masked input bit $a'$ is either \texttt{T} or \texttt{F}, i.e., it extracts the mask from the masked value. Using this, one can calculate \texttt{CountOnes} on a masked bitvector $\vec{a}'$ as
\begin{equation}
    \texttt{CountOnes}(\vec{a}') = \tfrac{1}{2}\cdot\texttt{Sum}(\vec{a}'\odot\vec{a}' - \vec{a}')\,,
\end{equation}
where $\odot$ is an element-wise multiplication.
Looking closer, the subtraction of $\vec{a}'$ from $\vec{a}'\odot \vec{a}'$ removes all \texttt{F} states, while adding the \texttt{T} states a second time, which gets normalized by the multiplication with $1/2$.

Consequently, the hamming distance $\texttt{CountOnes}((\vec{a} \oplus \vec{b}) \wedge \vec{m}_a \wedge \vec{m}_b)$ can be calculated as

\begin{align}
    \texttt{CountOnes}(\vec{a}'\odot\vec{b}') & = \tfrac{1}{2}\cdot\texttt{Sum}(\vec{a}'\odot\vec{a}' \odot\vec{b}'\odot\vec{b}' - \vec{a}' \odot \vec{b}')               \notag        \\
                                              & =\tfrac{1}{2}\cdot(\texttt{Sum}(\vec{a}'\odot\vec{a}' \odot\vec{b}'\odot\vec{b}') - \texttt{Sum}(\vec{a}' \odot \vec{b}'))       \notag \\
                                              & = \tfrac{1}{2}\cdot(\texttt{CountOnes}(\vec{m}_a\wedge\vec{m_b}) - \sum_i a'_i \cdot b'_i)\,.
\end{align}

Since one can use $\vec{a}' \odot \vec{a}'$ to extract the mask $\vec{m}_a$ as bitvector, and since an AND gate between bits can be represented by a simple multiplication, $\texttt{Sum}(\vec{a}' \odot \vec{a}' \odot \vec{b}' \odot \vec{b}')$ reduces to $\texttt{CountOnes}(\vec{m}_a\wedge\vec{m_b})$.
Consequently, using masked bitvectors allows us to remove two sums from the hamming distance calculation in \Cref{eq:hd}.

To simplify the comparison operation $\texttt{hd} < \texttt{MATCH\_RATIO} \cdot \texttt{ml}$ from \Cref{alg:iris_plain}, where $\texttt{hd} = \texttt{CountOnes}((\vec{c} \oplus C_\texttt{DB}[i]) \wedge \vec{m} \wedge M_\texttt{DB}[i])$ and $\texttt{ml} = \texttt{CountOnes}(\vec{m} \wedge M_\texttt{DB}[i])$, one can rewrite it using the masked bit representation:
\begin{align*}
    \texttt{hd}                                                                                                  & < \texttt{MATCH\_RATIO} \cdot \texttt{ml}                 \\
    \Leftrightarrow \texttt{CountOnes}((\vec{c} \oplus C_\texttt{DB}[i]) \wedge \vec{m} \wedge M_\texttt{DB}[i]) & < \texttt{MATCH\_RATIO} \cdot \texttt{ml}                 \\
    \Leftrightarrow  \frac{1}{2}\cdot(\texttt{ml} - \sum_j c'_j \cdot C'_\texttt{DB}[i]_j)                       & < \texttt{MATCH\_RATIO} \cdot \texttt{ml}                 \\
    \Leftrightarrow \langle c', C'_\texttt{DB}[i]\rangle                                                         & > (1 - 2 \cdot  \texttt{MATCH\_RATIO}) \cdot \texttt{ml}.
\end{align*}

\section{Introduction on ABY3} \label{ap:aby3}

In ABY3, arithmetic values $x \in \Z_{2^k}$ are shared additively, such that $\texttt{Share}(x) = \shared{x} = (x_1, x_2, x_3)$ and $x = \sum x_i$.
Then each party $i$ gets as its share the values $(x_i, x_{i-1})$ (wrapping using $(i \mod 3) + 1$).
Since additive sharing is used, linear operations, such as addition, subtraction, and multiplications with constants, can be performed on the shares without the parties having to communicate with each other.

Multiplications $\shared{z} = \shared{x} \cdot \shared{y}$, on the other hand, can not be computed purely without communication.
However, since each party has 2 additive shares, they can compute an additive share of the result without communication.
Namely
\begin{equation}
    z_i = x_i \cdot y_i + x_{i-1} \cdot y_i + x_i \cdot y_{i-1} + r_i\,,
\end{equation}
where $r_i$ is a freshly generated random share of 0 required for re-randomization which can be produced without communication in ABY3~\cite{DBLP:conf/ccs/MohasselR18}.
To translate the additive share $z_i$ to a replicated share $(z_i, z_{i-1})$ it suffices that party $i$ sends its share to party $i+1$.

When computing a dot-product $\shared{z} = \langle\shared{\vec{x}}, \shared{\vec{y}}\rangle = \sum \shared{x_i} \cdot \shared{y_i}$, one does not have to re-establish the replication after each individual multiplication $\shared{x_i} \cdot \shared{y_i}$, but can perform the summation on the additive shares and finally re-share the result.
Consequently, a dot product requires the same amount of communication as a multiplication in ABY3, i.e., it is independent of the length of the vectors $|\shared{\vec{x}}| = |\shared{\vec{y}}|$.

Observe that the multiplication can be rewritten as
\begin{align}
    z_i & = x_i \cdot y_i + x_{i-1} \cdot y_i + x_i \cdot y_{i-1} + r_i                   \\
        & = (x_i + x_{i-1}) \cdot (y_i + y_{i-1}) - x_{i-1} \cdot y_{i-1} + r_i\,. \notag
\end{align}
Thus, since we directly use the input shares in dot-products, we can do the following optimization.
First, one can pre-process the share $(x_i, x_{i-1})$ as $(x'_i, x'_{i-1}) = (x_i + x_{i-1}, x_{i-1})$.
Then, multiplication reduces to two plain multiplications plus re-randomization:

\begin{align}
    z_i & = x_i \cdot y_i + x_{i-1} \cdot y_i + x_i \cdot y_{i-1} + r_i \\
        & = x'_i \cdot y'_i - x'_{i-1} \cdot y'_{i-1} + r_i\,. \notag
\end{align}
Consequently, a dot-product of shared values reduces to two plain dot-products and resharing including re-randomization.

\section{Missing Algorithms} \label{ap:algorithms}

\begin{algorithm}
    \caption{$\mathsf{InpLocal}(x, i, 2^k) \to ([x]_{2^k})$.}
    \label{alg:InpLocal}
    \begin{algorithmic}
        \Require $x$ is known to parties $i,i+1$
        \State Parties $i,i+1$ interpret $x$ as an element of $\Z_{2^k}$.
        \State Party $i$ defines its replicated shares as $x_i = x$ and $x_{i-1} = 0$.
        \State Party $i + 1$ defines its replicated shares as $x_{i+1} = 0$ and $x_i = x$.
        \State Party $i + 2$ defines its replicated shares as $x_{i+2} = 0$ and $x_{i+1} = 0$.
        \State \Return $[x]_{2^k}$
    \end{algorithmic}
\end{algorithm}

\begin{algorithm}
    \caption{$\mathsf{ShareSplit}([x]_{2^k}) \to ([x_1[i]]_2, [x_2[i]]_2, [x_3[i]]_2)_{i \in[0,k)}$.}
    \label{alg:sharesplit}
    \begin{algorithmic}
        \State Parties $i,i+1$ interpret $x_i$ as an element of $(\Z_2)^k$, i.e., as a vector of $k$ bits.
        \State  $[x_1[j]]_2 \gets \mathsf{InpLocal}(x_1[j], 1, 2)$ for all $j \in [0,k)$.
        \State  $[x_2[j]]_2 \gets \mathsf{InpLocal}(x_2[j], 2, 2)$ for all $j \in [0,k)$.
        \State  $[x_3[j]]_2 \gets \mathsf{InpLocal}(x_3[j], 3, 2)$ for all $j \in [0,k)$.
        \State \Return $([x_1[i]]_2, [x_2[i]]_2, [x_3[i]]_2)_{i \in[0,k)}$
    \end{algorithmic}
\end{algorithm}

\begin{algorithm}
    \caption{$\mathsf{FA}([a]_2, [b]_2, [c]_2) \to ([s]_2, [o]_2)$. This algorithm computes a full adder with inputs $a,b$, carry input $c$, carry output $o$ and result $s$.}
    \label{alg:fa}
    \begin{algorithmic}
        \State $[t_1]_2 \gets [a]_2 + [c]_2$
        \State $[t_2]_2 \gets [b]_2 + [c]_2$
        \State $[s]_2 \gets [t_1]_2 + [b]_2$
        \State $[o]_2 \gets [t_1]_2 \cdot [t_2]_2 + [c]_2$
        \State \Return $([s]_2, [o]_2)$
    \end{algorithmic}
\end{algorithm}

\begin{algorithm}
    \caption{$\mathsf{BinAdd}(([a[m-1]]_2, \dots, [a[0]]_2), ([b[m-1]]_2, \dots, [b[0]]_2)) \to ([c[m]]_2, \dots, [c[0]]_2)$. Binary addition of two values (ripple-carry adder).}
    \label{alg:BinAdd_RCA}
    \begin{algorithmic}
        \State $[s[0]]_2 = [a[0]]_2 + [b[0]]_2$
        \State $[c]_2 = [a[0]]_2 \cdot [b[0]]_2$
        \For{$i$ in $1..m$}
        \State $([c]_2, [s[i]]_2) \gets \mathsf{FA}([a[i]]_2, [b[i]]_2, [c[i]]_2)$
        \EndFor
        \State \Return $([c]_2, [s[m - 1]]_2, \dots, [s[0]]_2)$
    \end{algorithmic}
\end{algorithm}

\begin{algorithm}
    \caption{$\mathsf{BitExtract}([x]_{2^k}, I) \to ([x[i]]_2)_{i \in I}$. This algorithm extracts the bits with indices in $I$ from the shared value $x$.}
    \label{alg:BitExtract}
    \begin{algorithmic}
        \State  $([x_1[i]]_2, [x_2[i]]_2, [x_3[i]]_2)_{i \in[0,k)} \gets \mathsf{ShareSplit}([x]_{2^k})$.
        \State $m \gets \max i \in I$
        \State  $([c[i]]_2, [s[i]]_2) \gets \mathsf{FA}([x_1[i]]_2, [x_2[i]]_2, [x_3[i]]_2)$ for all $i \in [0,m]$.
        \State  $([r[m]]_2, \dots [r[0]]_2) \gets \mathsf{BinAdd}(([c[m-1]]_2, \dots, [c[0]]_2, 0), ([s[m]]_2, \dots, [s[0]]_2))$.
        \State \Return $([r[i]]_2)_{i \in I}$
    \end{algorithmic}
\end{algorithm}

\begin{algorithm}
    \caption{$\mathsf{ConstLift}([a]_{c}, d) \to [ac]_{c\cdot d}$.}
    \label{alg:constlift}
    \begin{algorithmic}
        \State Party $i$ interprets $a_i$ as an element of $\Z_{c\cdot d}$ and computes $b_i = a_i \cdot d$.
        \State \Return $[b]_{c\cdot d}$
    \end{algorithmic}
\end{algorithm}

\begin{algorithm}
    \caption{$\mathsf{Lift}([x]_{2^k}, 2^{k+m}) \to [x]_{2^{k+m}}$.}
    \label{alg:Lift}
    \begin{algorithmic}
        \State Parties $i,i+1$ interpret $x_i$ as an element of $\Z_{2^{k+m}}$
        \State $[x_1]_{2^{k+m}} \gets \mathsf{InpLocal}(x_1, 1, 2^{m+k})$.
        \State $[x_2]_{2^{k+m}} \gets \mathsf{InpLocal}(x_2, 2, 2^{m+k})$.
        \State $[x_3]_{2^{k+m}} \gets \mathsf{InpLocal}(x_3, 3, 2^{m+k})$.
        \State $[x_1+x_2+x_3]_{2^{k+m}} \gets [x_1]_{2^{k+m}} + [x_2]_{2^{k+m}} + [x_3]_{2^{k+m}}$.
        \State $([a]_2,[b]_2) \gets \mathsf{BitExtract}([x_1+x_2+x_3]_{2^{k+m}}, \{k+1,k\})$.
        \State $[a]_{2^{m-1}} \gets \mathsf{BitInject}([a]_2, 2^{m-1})$.
        \State $[b]_{2^m} \gets \mathsf{BitInject}([b]_2, 2^{m})$.
        \State $[2^{k+1}a]_{2^{k+m}} \gets \mathsf{ConstLift}([a]_{2^{m-1}}, 2^{k+1})$.
        \State $[2^{k}b]_{2^{k+m}} \gets \mathsf{ConstLift}([b]_{2^m}, 2^{k})$.
        \State \Return $[x_1 + x_2 + x_3]_{2^{k+m}} - [2^{k+1}a]_{2^{k+m}} - [2^{k}b]_{2^{k+m}}$
    \end{algorithmic}
\end{algorithm}

\begin{algorithm}
    \caption{3-Party OT: $\mathsf{3OT}((m_0,m_1), c, c) \to (\bot, m_c, \bot)$. (Party 1 is the OT sender, Party 2 is the OT receiver, Party 3 is the OT helper. Let $m_i \in \Z_{2^k}$.) Based on the 3-party OT in \cite{DBLP:conf/ccs/MohasselR18}.}
    \label{alg:3ot}
    \begin{algorithmic}
        \Require Party $i$ holds $\mathsf{seed}_i, \mathsf{seed}_{i-1}$
        \State Parties $1,3$ generate $w_0, w_1 \sample \mathsf{RNG.Gen}(\mathsf{seed}_3, k)$
        \State Party $1$: $k_0 \gets w_0 \oplus m_0$, $k_1 \gets w_1 \oplus m_1$
        \State Party $1$ sends $k_0, k_1$ to Party $2$
        \State Party $3$ sends $w_c$ to Party $2$
        \State Party $2$: $m_c \gets w_c \oplus k_c$ and outputs $m_c$
    \end{algorithmic}
\end{algorithm}

\begin{algorithm}
    \caption{$\mathsf{BitInject}([x]_{2}, 2^k) \to [x]_{2^k}$. Based on \cite{DBLP:conf/ccs/MohasselR18}.}
    \label{alg:bitinject}
    \begin{algorithmic}
        \Require Party $i$ holds $\mathsf{seed}_i, \mathsf{seed}_{i-1}$
        \State Parties $1,2$ generate $c_1 \sample \mathsf{RNG.Gen}(\mathsf{seed}_1, k)$
        \State Parties $1,3$ generate $c_3 \sample \mathsf{RNG.Gen}(\mathsf{seed}_3, k)$
        \State Party 1: $m_i \gets (i \oplus x_1 \oplus x_3) - c_1 - c_3$ for $i \in \{0,1\}$.
        \State Execute $\mathsf{3OT}((m_0, m_1), x_2, x_2) \to (\bot, m_{x_2}, \bot)$
        \State Party 2 sets $c_2 \gets m_{x_2}$ and sends $c_2$ to Party 3.
        \State \Return $[c]_{2^k}$
    \end{algorithmic}
\end{algorithm}

\end{document}